\documentclass[reprint,showpacs,amsmath,amssymb,aps]{revtex4-1}

\usepackage{graphicx}
\usepackage{dcolumn}
\usepackage{bm}
\allowdisplaybreaks

\usepackage{mathrsfs}
\usepackage{mathtools}

\begin{document}

\title{Percolation in Random Graphs:~A Finite Approach}

\author{Michelle Rudolph-Lilith}
\email[Electronic address: ]{rudolph@unic.cnrs-gif.fr}
\author{Lyle E. Muller}
\affiliation{
Unit\'e de Neurosciences, Information et Complexit\'e (UNIC) \\
CNRS, 1 Ave de la Terrasse, 91198 Gif-sur-Yvette, France}

\date{\today}

\begin{abstract}
We propose an approach to calculate the critical percolation threshold for finite-sized Erd\H{o}s-R\'enyi digraphs using minimal Hamiltonian cycles. We obtain an analytically exact result, valid non-asymptotically for all graph sizes, which scales in accordance with results obtained for infinite random graphs using the emergence of a giant connected component as marking the percolation transition. Our approach is general and can be applied to all graph models for which an algebraic formulation of the adjacency matrix is available.
\end{abstract}

\pacs{64.60.aq,64.60.ah,64.60.an}


\maketitle


The {\it exempla} that eventually became the inspiration for percolation theory in random graphs were originally defined in terms of paths \cite{HammersleyMorton54}. Consider for example a porous medium with a liquid flowing downwards along paths probabilistically connecting the top to bottom. Although intuitive, a mathematically rigorous treatment of such a problem quickly runs into a combinatorical explosion \cite{Broadbent64}. Thus, despite its practical importance, to this date analytically exact results remain sparse.

To overcome these problems, several simplifications were proposed in the study of percolation in random graphs, most notably the consideration of infinite system size (for reviews, see \cite{Hofstad10, Grimmett99}). In this limit, the search for spanning paths becomes meaningless, however, and is commonly abstracted by asking whether a significantly large connected cluster exists. The percolation transition is then identified with the appearance of a spanning giant connected component containing $\mathcal{O}(N_N)$ nodes, where $N_N \rightarrow \infty$ is the number of nodes in the graph.

Various analytical methods have been introduced to assess the size distribution of connected components in random graphs in the asymptotic limit. An approach using generating functions to assess the emergence of connected components of specific size assumes infinite but locally finite quasi-transitive graphs in which no closed paths, or cycles, exist \cite{Newman01}. Here, the search for the giant connected component is equivalent to the search for a spanning tree. Statistical \cite{AlbertBarabasi02} or mean-field approaches \cite{HaraSlade90} generally provide only scaling results in the asymptotic limit. Independent of the methods utilized, however, the percolation threshold, i.e.~the critical connection probability at which a giant connected component appears, was found to be $p_c \simeq 1/N_N$ in random graphs, with the size of the giant component scaling with $\ln(N_N)$ below $p_c$ and $N_N^{2/3}$ at $p_c$ \cite{Hofstad10,Grimmett99,AlbertBarabasi02}.

Here, we follow an operator graph-theoretic method introduced in \cite{RudolphLilithMuller14}, which allows to calculate algebraically well-defined graph measures exactly in the non-asymptotic limit. We propose an approach to percolation in finite random graphs based on closed paths, or cycles. This notion is not only closer to the original conception of the percolation phenomenon, but also mathematically tractable in finite directed Erd\H{o}s-R\'enyi random graphs. Specifically, we define the percolation threshold $p_c$ as the critical connection probability at which the first minimal Hamiltonian cycle of length $N_N$, defined as a closed walk that visits each node exactly once, occurs. We note that the occurrence of this minimal Hamiltonian cycle is a sufficient condition for the emergence of a giant connected component spanning the full graph. 

We start by constructing a non-self-looped Erd\H{o}s-R\'enyi digraph algebraically. To that end, we introduce a binomial random annihilation operator, defined as
\begin{equation}
\label{Eq_Rp}
\hat{r}^p(x) = \left\{ \begin{array}{ll}
x & \text{with probability } p \\
0 & \text{with probability } 1-p 
\end{array} \right.
\end{equation}
and understood statistically, i.e. the sum over $n$ applications of $\hat{r}^p$ on $x$ returns $\sum^{n} \hat{r}^p(x) = npx$. It can easily be demonstrated that the set of these operators form a linear algebra which is both commutative and associative under multiplication, as well as distributive. Furthermore, we introduce an $N_N \times N_N$ circulant matrix
\begin{equation}
\label{Eq_One}
\boldsymbol{1}_{ij}
= \text{circ}\big( \{ 0, \overbrace{1, \ldots, 1}^{N_N-1} \} \big) 
= \text{circ}\bigg[ \Big( \sum\limits_{l=1}^{N_N-1} \delta_{1+l,j} \Big)_j \bigg] .
\end{equation}
With (\ref{Eq_Rp}) and (\ref{Eq_One}), the elements of the adjacency matrix $\boldsymbol{A}$ of a non-self-looped Erd\H{o}s-R\'enyi digraph with connectedness $p$ is given by
\begin{equation}
\label{Eq_AdjacencyMatrix}
a_{ij} = ( 1 - \hat{r}^{1-p}_{ij} ) \boldsymbol{1}_{ij} \, ,
\end{equation}
with $\hat{r}^{p}_{ij}$ denoting a matrix of independent random annihilation operators $\hat{r}^{p}$.

\begin{figure}[t]
\includegraphics[width=\columnwidth]{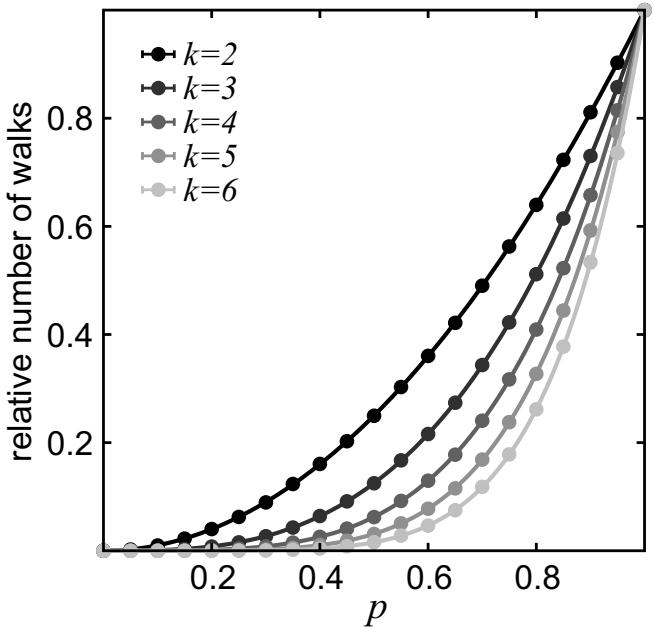}
\caption{\label{Fig_N0k} 
Relative number $\mathring{\mathcal{N}}_k / \mathring{\mathcal{N}}_k|_{p=1}$ of closed walks of length $k$ in a not self-looped Erd\H{o}s-R\'enyi digraph as function of connectedness $p$ for a graph of $N_N=100$ nodes. Shown are the numerical result (dots) and analytical solution [lines; Eq.~(\ref{Eq_N0k_solution})]. For a given $k$, the relative number of walks increases $\sim p^k$, independent of the size of the graph. For the numerical model, 100 random realizations were used for each parameter set. The error bars on the numerical results are smaller than the data points.}
\end{figure}

To demonstrate the application of Eq.~(\ref{Eq_AdjacencyMatrix}), we calculate the number $\mathring{\mathcal{N}}_k$ of closed walks of length $k$, defined as the trace over the $k$th power of an adjacency matrix,
\begin{equation}
\label{Eq_N0k}
\mathring{\mathcal{N}}_k 
= \text{Tr}( \boldsymbol{A}^k )_{ij} 
= p^k \text{Tr}( \boldsymbol{1}^k )_{ij} \, ,
\end{equation}
where in the last step the statistical nature of $\hat{r}^{p}_{ij}$ was employed. As $\boldsymbol{1}_{ij}$ is a circulant matrix, application of the circulant diagonalization theorem yields
\begin{equation*}
( \boldsymbol{1}^k )_{rs}
= \frac{(N_N-1)^k}{N_N} + \frac{(-1)^k}{N_N} \sum\limits_{l=1}^{N_N-1} 
  \exp\left[ -\frac{2 \pi i}{N_N} l (r-s) \right]
\end{equation*}
for its $k$th power. Inserting the latter into (\ref{Eq_N0k}), one obtains for the total number of closed walks of length $k$ in a non-self-looped Erd\H{o}s-R\'enyi digraph with connectedness $p$ and size $N_N$
\begin{equation}
\label{Eq_N0k_solution}
\mathring{\mathcal{N}}_k = p^k \big[ (N_N-1)^k + (-1)^k (N_N-1) \big] .
\end{equation}
Figure~\ref{Fig_N0k} compares the numerical result and corresponding analytical solution for a small graph of $N_N=100$ nodes.

Next we consider general walks of length $k$ visiting $n$ distinct nodes. To that end, let $I^k = \{ i_1, \ldots, i_k \}$ with $|I^k|=k$ being a set of indices. We consider labelled partitions of this set into two unordered subsets $I_1^n$ and $I_2^{k-n}$ with $|I_1^n|=n$ and $|I_2^{k-n}|=k-n$, respectively, such that
\begin{eqnarray}
I^{k|n} & = & I_1^n \cup I_2^{k-n} \nonumber \\
I_1^n \cap I_2^{k-n} & = & \varnothing \, .
\end{eqnarray}
Denoting by $\{ I^{k|n} \}$ the set of all such partitions, we can define a generalized 
\begin{equation}
\label{Eq_DeltaKN}
\delta_{k|n}
= \left\{ 
\begin{array}{lll}
\hspace*{2mm}  & \displaystyle\sum\limits_{\mathclap{ \{ I^{k+1|n} \}} } \, \bar{\delta}_{I_1^n} \equiv \bar{\delta}_{I^{k+1}} & n=k+1 \\
  & \displaystyle\sum\limits_{\mathclap{ \substack{ (I_1^{n-1} \cup I_2^{k-n+2}) \\ \in \{ I^{k+1|n-1} \} } } } \, \bar{\delta}_{I_1^{n-1}} \delta_{I_2^{k-n+2}} \bar{\delta}_{I_1^{n-1}, I_2^{k-n+2}} & 1 \leq n \leq k ,
\end{array}
\right.
\end{equation}
where for given index sets $\mathcal{A}$ and $\mathcal{B}$
\begin{eqnarray}
\label{Eq_Delta}
\delta_{\mathcal{A}} & = & \prod\limits_{(i,j) \in \mathcal{P}_{\mathcal{A}}} \delta_{ij} \nonumber \\
\bar{\delta}_{\mathcal{A}} & = & \prod\limits_{(i,j) \in \mathcal{P}_{\mathcal{A}}} (1 - \delta_{ij}) \nonumber \\
\bar{\delta}_{\mathcal{A},\mathcal{B}} & = & \prod\limits_{(i,j) \in \mathcal{A} \times \mathcal{B}} (1 - \delta_{ij} ) . 
\end{eqnarray}
Here, $\mathcal{P}_{\mathcal{A}}$ denotes the set of all unordered pairs $(i,j)$ with $i,j \in \mathcal{A}$, and $\delta_{ij}$ the Kronecker delta. Given two index sets $I_1^l$ and $I_2^{l'}$, Eqs.~(\ref{Eq_DeltaKN}) and (\ref{Eq_Delta}) algebraically formulate that all indices in $I_1^l$ are mutually distinct, all indices in $I_2^{l'}$ are mutually equal, and each index from $I_1^l$ is distinct from each index in $I_2^{l'}$.

Generalizing Eq.~(\ref{Eq_N0k}), the number of walks and closed walks, $\mathcal{N}_k^n$ and $\mathring{\mathcal{N}}_k^n$, respectively, of length $k$ visiting $n$ distinct nodes is given by the $k$th power of the graph's adjacency matrix, with restrictions imposed on the indices to ensure that only $n$ distinct nodes are visited. With (\ref{Eq_DeltaKN}), we have
\begin{eqnarray}
\mathcal{N}_k^n
& = & \sum\limits_{i_1, \ldots, i_{k+1} = 1}^{N_N} \delta_{k|n} \prod\limits_{l=1}^{k} a_{i_l i_{l+1}} \label{Eq_Nnk} \\ 
\mathring{\mathcal{N}}_k^n
& = & \sum\limits_{i_1, \ldots, i_k = 1}^{N_N} \delta_{k-1|n} \left( \prod\limits_{l=1}^{k-1} a_{i_l i_{l+1}} \right) a_{i_k i_1} \label{Eq_N0nk} \, . 
\end{eqnarray}
We note that the latter is constructed from open walks of length $k-1$ visiting $n$ nodes by adding one more edge connecting the last node in the walk with its first node.

Restricting to the special case $n=k+1$, an analytically closed form for $\mathcal{N}_k^n$ can be obtained by observing the recursion
\begin{equation}
\label{Eq_deltaAlg}
\bar{\delta}_{I^{k+1}} = \bar{\delta}_{I^k} \prod\limits_{l=1}^{k} ( 1 - \delta_{i_l i_{k+1}} ) \, ,
\end{equation}
which can easily be shown using set-theoretical considerations. Inserting (\ref{Eq_AdjacencyMatrix}) into (\ref{Eq_Nnk}) and using (\ref{Eq_deltaAlg}), we obtain
\begin{equation}
\mathcal{N}_k^{k+1} 
= p (N_N - k) \mathcal{N}_{k-1}^k 
= p^{k-1} \frac{\Gamma[N_N-1]}{\Gamma[N_N-k]} \, \mathcal{N}_1^2 \, .
\end{equation}
The term $\mathcal{N}_1^2$ denotes the number of walks of length 1 visiting 2 distinct nodes, which, in a non-self-looped digraph, is equivalent to the graph's total adjacency $A$, thus yielding finally
\begin{equation}
\mathcal{N}_k^{k+1} = p^{k-1} \frac{\Gamma[N_N-1]}{\Gamma[N_N-k]} \, A
\end{equation}
for the total number of walks of length $k$ visiting $k+1$ distinct nodes in an Erd\H{o}s-R\'enyi digraph with connectedness $p$ and size $N_N$. 

\begin{figure}[t]
\includegraphics[width=\columnwidth]{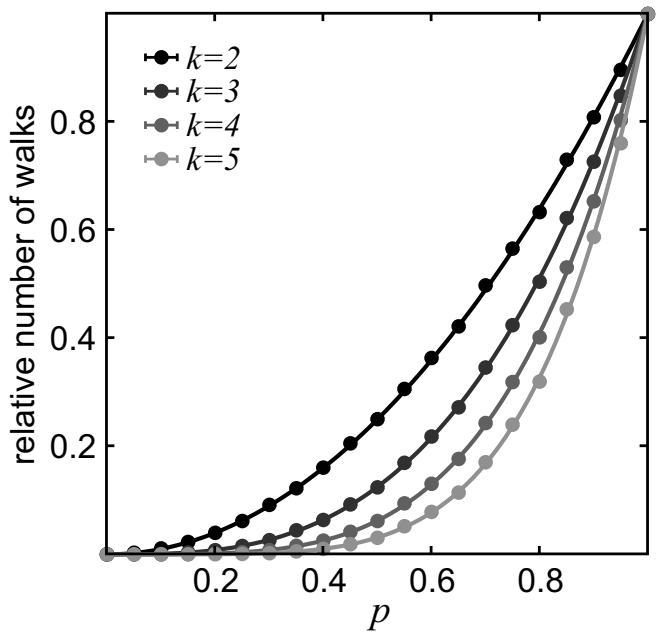}
\caption{\label{Fig_N0kk} 
Relative number $\mathring{\mathcal{N}}_k^k / \mathring{\mathcal{N}}_k^k|_{p=1}$ of closed walks of length $k$ visiting $k$ nodes in a not self-looped Erd\H{o}s-R\'enyi digraph as function of connectedness $p$, for a graph of $N_N=19$ nodes. Shown are the numerical result (dots) and analytical solution [lines; Eq.~(\ref{Eq_N0kk})]. For a given $k$, the relative number of walks increases $\sim p^{k-1}$, independent of the size of the graph. For the numerical model, 100 random realizations were used for each parameter set.}
\end{figure}

Similarly, inserting (\ref{Eq_AdjacencyMatrix}) into (\ref{Eq_N0nk}), we obtain
\begin{eqnarray}
\mathring{\mathcal{N}}_k^k
& = & p \mathcal{N}^k_{k-1} 
  =   p (N_N - k + 1) \mathring{\mathcal{N}}_{k-1}^{k-1} \nonumber \\
& = & p^{k-2} \frac{\Gamma[N_N-1]}{\Gamma[N_N-k+1]} \, \mathring{\mathcal{N}}_2^2 \, ,
\end{eqnarray}
where $\mathring{\mathcal{N}}_2^2$ denotes the number of closed walks of length 2 visiting 2 distinct nodes. The latter is equivalent to twice the number of bidirectional connected node pairs in non-self-looped random digraphs, $pA/2$, thus yielding
\begin{equation}
\label{Eq_N0kk}
\mathring{\mathcal{N}}_k^k = p^{k-1} \frac{\Gamma[N_N-1]}{\Gamma[N_N-k+1]} \, A
\end{equation}
for the total number of closed walks of length $k$ visiting $k$ distinct nodes. Figure~\ref{Fig_N0kk} compares the numerical result and corresponding analytical solution for a small graph of $N_N=19$ nodes. We note that the number of nodes in the numerical analysis was kept small as the search for specific walks constitutes an NP-hard problem, and requires significant computational resources for larger graphs. 

With Eq.~(\ref{Eq_N0kk}), we can now proceed to address the percolation threshold in terms of minimal Hamiltonian cycles, given by $\mathring{\mathcal{N}}_{N_N}^{N_N}$. Percolation is here defined to occur when there is at least one such cycle. Due to the symmetry of cycles of length $N_N$, if one such cycle emerges, there are $N_N$ such cycles present in the graph. Thus, the critical connectedness $p_c$ can be defined as the connectedness for which $\mathring{\mathcal{N}}_{N_N}^{N_N} = N_N$. With (\ref{Eq_N0kk}), we obtain
$$
N_N = p_c^{N_N-1} \frac{\Gamma[N_N-1]}{\Gamma[1]} \, A \, ,
$$
which yields, with $A = p N_N (N_N-1)$ for random graphs,
\begin{equation}
\label{Eq_CriticalConnectedness}
p_c = \big( \Gamma[N_N] \big)^{-\frac{1}{N_N}}
\end{equation}
for the critical percolation threshold for Erd\H{o}s-R\'enyi digraph of size $N_N$.

\begin{figure}[t]
\includegraphics[width=\columnwidth]{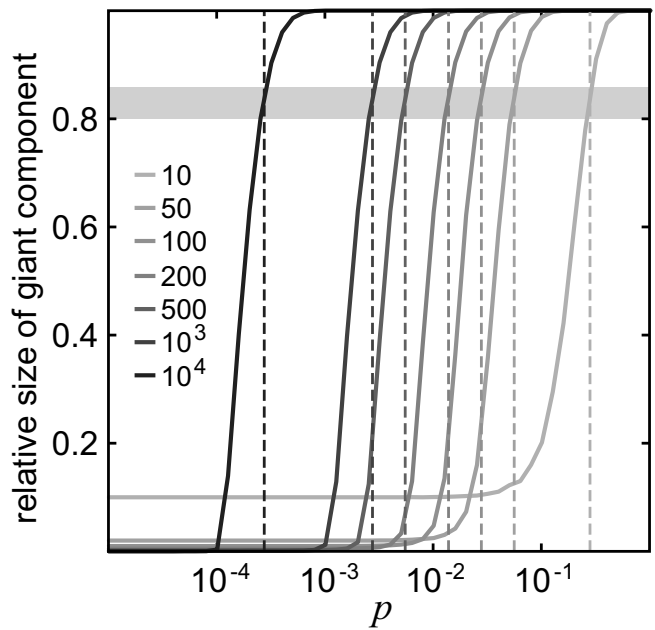}
\caption{\label{Fig_Sgcc} 
Relative size of the giant connected component as function of the connectivity $p$ in not self-looped Erd\H{o}s-R\'enyi digraph of various size $N_N$. Solid lines show the numerical average over 1,000 random realizations for each parameter set (100 for $N_N = 10,000$), dashed lines indicate the critical percolation threshold $p_c$, Eq.~(\ref{Eq_CriticalConnectedness}).}
\end{figure}

In order to compare the result (\ref{Eq_CriticalConnectedness}) with the emergence of a dominant giant connected component, which is most commonly used to characterize the percolation transition, we numerically generated Erd\H{o}s-R\'enyi digraphs of various size and connectedness, and investigated the average size of their respective giant connected components (Fig.~\ref{Fig_Sgcc}, solid). The critical threshold $p_c$ [Eq.~(\ref{Eq_CriticalConnectedness}); Fig.~\ref{Fig_Sgcc}, dashed] lies within the sharp percolation transition, marked by the emergence of a dominant giant component. Moreover, the numerical analysis indicates that $p_c$ consistently coincides with the emergence of a giant component covering about 80-85\% of the graph (Fig.~\ref{Fig_Sgcc}, gray bar), independent of the graph size within the investigated parameter regime. This finding suggests that, at percolation threshold $p_c$, the size of the giant component will scale linearly with the graph size $N_N$. 

We note that the latter stands in stark contrast to the classical result, which finds a scaling with $N_N^{2/3}$ \cite{Hofstad10, Grimmett99, AlbertBarabasi02}. However, this classical result, which uses the emergence of a giant component as marking the percolation transition, must be viewed with care, as it yields a relative size of the giant connected component which scales as $N_N^{-1/3} \rightarrow 0$ for $N_N \rightarrow \infty$. Thus, for infinite graphs, the giant component would occupy an infinitesimal fraction of the whole graph and not $\mathcal{O}(1)$, as required.

Finally, we investigated the asymptotical behavior of $p_c$. Using Stirling's approximation, Eq.~(\ref{Eq_CriticalConnectedness}) yields
\begin{align}
p_c\big|_{N_N \rightarrow \infty}
&= e^{\frac{N_N-1}{N_N}} (2 \pi)^{-\frac{1}{2 N_N}} (N_N - 1)^{-\frac{2 N_N - 1}{2 N_N}} \nonumber \\  
&\sim \frac{1}{N_N} \, .
\end{align}
This corresponds to the classical asymptotic scaling result for the percolation threshold in infinite random graphs \cite{Hofstad10, Grimmett99, AlbertBarabasi02}. 

In this paper, we have investigated the percolation transition for finite-size simple random digraphs in a context close to its original conception \cite{HammersleyMorton54}, defined as the first occurrence of a path, or walk, spanning the whole system. To that end, we have calculated the expected total number of closed walks of length $k$ [$\mathring{\mathcal{N}}_k$; Eq.~(\ref{Eq_N0k_solution})] and total number of closed walks of length $k$ visiting $k$ distinct nodes [$\mathring{\mathcal{N}}_k^k$; Eq.~(\ref{Eq_N0kk})] in Erd\H{o}s-R\'enyi digraphs of connectedness $p$ and size $N_N$. The latter expression was then used to calculate the critical connectedness at which the first minimal Hamiltonian cycle emerges, thus quantifying non-asymptotically and analytically exact the percolation threshold $p_c$.

In contrast to the classical definition of percolation in random graphs, which is meaningful only for infinite systems and uses the emergence of the giant component of size $\mathcal{O}(N_N)$ to mark the percolation transition, walks on graphs are an algebraically well-defined quantity and can be calculated exactly in cases where an explicit algebraic form of the adjacency matrix of the graph is available. Our approach is general and can be applied to characterize the percolation transition in other graph models for which an algebraic formulation of the adjacency matrix is available. A mathematically rigorous presentation of this framework is in active development.


\begin{acknowledgments}
The authors wish to thank OD Little for comments. This work was supported by CNRS, the European Community (BrainScales Project No. FP7-269921), and \'Ecole des Neurosciences de Paris Ile-de-France.
\end{acknowledgments}



\begin{thebibliography}{9} 

\bibitem{HammersleyMorton54} J.M. Hammersley, K.W. Morton, J. Roy. Stat. Soc. B {\bf 16}, 23 (1954).

\bibitem{Broadbent64} S.R. Broadbent, J. Roy. Stat. Soc. B {\bf 16}, 68 (1964).

\bibitem{Hofstad10} R. van der Hofstad, In: {\it New Perspectives on Stochastic Geometry} (Oxford Univ. Press, 2010).

\bibitem{Grimmett99} G.R. Grimmett, {\it Percolation} (Springer, 1999).

\bibitem{Newman01} M.E.J. Newman {\it et al.}, Phys. Rev. E {\bf 64}, 026118 (2001).

\bibitem{AlbertBarabasi02} R. Albert, A.-L. Barab\'asi, Rev. Mod. Phys. {\bf 74}, 47 (2002).

\bibitem{HaraSlade90} T. Hara, G. Slade, Comm. Math. Phys. {\bf 128}, 333 (1990).

\bibitem{RudolphLilithMuller14} M. Rudolph-Lilith, L.E. Muller, Phys. Rev. E {\bf 89}, 012812 (2014).

\end{thebibliography}
\end{document}